\documentclass[aps,preprint,showpacs,preprintnumbers,amsmath,amssymb]{revtex4}

\usepackage{amsmath,mathrsfs,amsbsy,color,graphicx,bm,amsthm,amsfonts}
\usepackage{units}
\usepackage{bbm}
\usepackage{times}
\usepackage{dcolumn}
\usepackage{mathrsfs}
\usepackage{amsmath,amssymb,epsfig}
\usepackage{amsmath,amssymb,amsfonts}
\usepackage{mathrsfs,amsbsy}
\usepackage{color,graphicx}
\usepackage{bm,amsthm}
\usepackage{units}
\usepackage{bbm}
\usepackage{times}
\usepackage{dcolumn}
\usepackage{epsfig}
\usepackage{amsmath, amssymb}
\usepackage{amsmath}
\usepackage{float} 
\usepackage{caption}
\newcommand{\udots}{\mathinner{\mskip1mu\raise1pt\vbox{\kern7pt\hbox{.}}
\mskip2mu\raise4pt\hbox{.}\mskip2mu\raise7pt\hbox{.}\mskip1mu}}

\begin{document}

\title{Directional effects of the Unruh effect on  steering harvesting}
\author{Shu-Min Wu$^1$ \footnote{smwu@lnnu.edu.cn }, Rui-Yang Xu$^1$,  Zhihong Liu$^2$ \footnote{zhihongliu@hunnu.edu.cn (corresponding author)} }
\affiliation{$^1$  Department of Physics, Liaoning Normal University, Dalian 116029, China \\
$^2$  Department of Physics, Changzhi University, Changzhi, 046011, China
}


\begin{abstract}
We investigate directional quantum steering harvesting by two uniformly accelerated non-identical Unruh-DeWitt detectors coupled to a massless scalar field in the Minkowski vacuum. We show that the Unruh effect does not simply degrade harvested steering, but can reshape its directionality in a strongly parameter-dependent manner. At low detector energy scales, acceleration predominantly suppresses steering, whereas at higher energy scales, moderate acceleration can enhance steering and even activate an initially absent steering direction. Most strikingly, within a finite acceleration window, the Unruh effect enhances steering in one direction while simultaneously suppressing it in the opposite direction, revealing an intrinsically directional influence of relativistic acceleration. Detector non-identicality provides an additional control mechanism: a moderate energy-gap mismatch can enhance or activate one steering direction while suppressing the reverse one, whereas sufficiently large energy-gap mismatch suppresses the harvested steering. Moreover, the maximal steering asymmetry approximately marks the transition from two-way to one-way steering. Our results establish relativistic acceleration and detector non-identicality as means of controlling the strength and directionality of vacuum-harvested quantum correlations.
\end{abstract}

\vspace*{0.5cm}
\maketitle

\section{Introduction}
Quantum steering is an asymmetric form of nonlocal quantum correlation, whereby local measurements performed by one observer can remotely steer the state of another spatially separated subsystem \cite{M1}. Originally introduced by Schr\"odinger in response to the Einstein-Podolsky-Rosen (EPR) paradox \cite{M2,M3}, quantum steering was later rigorously formulated by Wiseman \textit{et al.} within the framework of local-hidden-state models \cite{M4}. In the hierarchy of quantum correlations, steering occupies an intermediate position between entanglement and Bell nonlocality \cite{M5}. A distinctive feature of quantum steering is its intrinsic directionality: the ability of subsystem $A$ to steer subsystem $B$ does not necessarily imply that $B$ can steer $A$. This directional asymmetry can lead to one-way steering, where steering exists in one direction but is absent in the reverse direction \cite{M6,M7}. Such sensitivity to asymmetries between subsystems makes quantum steering a useful probe of directional quantum correlations and an important resource for one-sided device-independent quantum-information protocols, including quantum communication, quantum metrology, and quantum networks \cite{M9,M10,M11,M12}.

Relativistic quantum information explores how relativistic motion and spacetime structure affect the generation, transmission, and manipulation of quantum information \cite{SDF1,SDF2,SDF3,SDF4,SDF5,SDF6,SDF7,SDF8,SDF9,SDF10,SDF11,SDF12,SDF13,SDF14,SDF15,SDF16,SDF17,SDF18,SDF19,SDF20,SDF21,SDF22,SDF23,SDF24,SDF25,SDF26,SDF27,SDF28,SDF29,SDF30,SDF31,SDF32,SDF33,SDF34,SDF35,SDF36,SDF37,SDF38,SDF39,SDF40,SDF41,SDF42,SDF43,SDF44,SDF45,SDF46,SDF47,SDF48}. Previous studies have shown that relativistic effects can degrade bipartite quantum steering and induce different decay rates in the two steering directions \cite{AGKL1,AGKL2,AGKL3}, thereby generating steering asymmetry. Most of these studies, however, focus on initially entangled systems, where relativistic effects modify pre-existing quantum correlations. A conceptually different scenario arises when initially uncorrelated systems locally interact with a quantum field and extract correlations from vacuum fluctuations. This possibility was first explored by Valentini and subsequently developed by Reznik, who showed that two initially uncorrelated detectors can harvest entanglement from the vacuum even when they remain spacelike separated \cite{V1,V2}. Since then, entanglement harvesting has been extensively investigated in different spacetime backgrounds and detector-motion scenarios, revealing a sensitive dependence on spacetime dimensionality, topology, curvature, causal structure, boundary conditions, and detector parameters such as trajectory, coupling strength, and energy gap \cite{A1,A2,A3,A4,A5,A6,A7,A8,A9,A10,A111,A131,A14,A15,A16,A17,A18,A20,B51,B52,B53,B54,QVP15}. Despite substantial progress in entanglement harvesting, the directional properties of harvested quantum steering in relativistic settings remain largely unexplored. In particular, it is not yet clear whether relativistic acceleration merely suppresses steering in both directions at different rates or can play a more intricate directional role, enhancing steering in one direction while simultaneously suppressing it in the opposite direction. This question is especially relevant to the Unruh effect, since acceleration modifies both the local detector response and the field-mediated correlations from which steering is harvested. Understanding this interplay is therefore important for clarifying how relativistic motion reshapes not only the amount of vacuum-harvested quantum correlations but also their directionality.

In this work, we investigate directional quantum steering harvesting from the vacuum using two non-identical Unruh-DeWitt detectors, labeled $A$ and $B$, undergoing uniform acceleration along the same spatial direction, with detector $B$ having a larger energy gap than detector $A$. We systematically examine how acceleration, detector separation, and energy-gap difference jointly affect the strength, asymmetry, and directional regimes of the harvested steering. In particular, we explore whether the Unruh effect merely suppresses steering or can instead produce direction-dependent responses, including the enhancement or activation of one steering direction accompanied by the suppression of the other. We further investigate how detector non-identicality modifies this directional behavior and drives transitions among two-way, one-way, and no-way steering regimes. Our results reveal that relativistic acceleration can play a dual and intrinsically directional role in steering harvesting, while the energy-gap difference provides an additional control parameter for manipulating the directionality of vacuum-harvested quantum correlations.

The remainder of this paper is organized as follows. In Sec.~II, we develop the physical model of two uniformly accelerated Unruh-DeWitt detectors with unequal energy gaps and derive their reduced density matrix. In Sec.~III, we formulate the criteria and measures for directional quantum steering and investigate how acceleration, detector separation, and energy-gap difference affect the strength, asymmetry, and directionality of the harvested steering. Finally, Sec.~IV summarizes our main results and conclusions.

\section{Non-identical accelerated Unruh-DeWitt detectors}

We consider two non-identical two-level Unruh-DeWitt detectors, labeled $A$ and $B$, with ground states $|0\rangle_D$, excited states $|1\rangle_D$, and energy gaps $\Omega_D$ ($D\in\{A,B\}$). Each detector interacts locally with a massless real scalar field $\phi[x_D(\tau)]$ in the Minkowski vacuum $|0\rangle_M$. In the interaction picture, the detector-field coupling is described by the  interaction Hamiltonian
\begin{equation}
H_D(\tau) = \lambda \chi(\tau) \left( e^{i\Omega_D\tau} \sigma_+ + e^{-i\Omega_D\tau} \sigma_- \right) \otimes \phi[x_D(\tau)], \quad D \in \{A, B\},
\label{1}
\end{equation}
where $\sigma_+=|1_D\rangle\langle0_D|$ and
$\sigma_-=|0_D\rangle\langle1_D|$ are the raising and lowering operators of detector $D$, respectively. The detector trajectory $x_D(\tau)$ is parametrized by its proper time $\tau$, $\lambda$ denotes the coupling strength, and  $\chi(\tau) = \exp[-\tau^2/(2\sigma^2)]$
is a Gaussian switching function, with $\sigma$ setting the characteristic interaction timescale.

We consider both detectors to be initially prepared in their ground states, while the field is in the Minkowski vacuum. The initial state of the detector-field system is therefore
\begin{equation}
|\Psi_i\rangle = |0\rangle_A |0\rangle_B |0\rangle_M.
\label{2}
\end{equation}
The final state is obtained from the time-ordered evolution
\begin{equation}
|\Psi_f\rangle = T \exp\left[-i \int dt \left( \frac{d\tau_A}{dt} H_A(\tau_A) + \frac{d\tau_B}{dt} H_B(\tau_B) \right) \right] |\Psi_i\rangle,
\label{3}
\end{equation}
where $T$ denotes the time-ordering operator and $t$ is the coordinate time with respect to which the field vacuum is defined. Expanding the time-evolution operator perturbatively and tracing out the field degrees of freedom, the reduced state of the two detectors, up to second order in the coupling strength $\lambda$, takes the $X$-state form \cite{V3,V4}
\begin{equation}
\rho_{AB} := \text{Tr}_\phi(|\Psi_f\rangle\langle\Psi_f|) =
\begin{pmatrix}
1 - P_A - P_B & 0 & 0 & X \\
0 & P_B & C & 0 \\
0 & C^* & P_A & 0 \\
X^* & 0 & 0 & 0
\end{pmatrix}
+ \mathcal{O}(\lambda^4),
\label{4}
\end{equation}
in the ordered basis
$\{|0_A0_B\rangle,|0_A1_B\rangle,|1_A0_B\rangle,|1_A1_B\rangle\}$. The matrix elements are given by
\begin{align}
P_D &:= \lambda^2 \iint d\tau d\tau' \chi(\tau)\chi(\tau') e^{-i\Omega_D(\tau-\tau')} W(x_D(\tau), x_D(\tau')), \quad D \in \{A, B\},\label{5} \\
C &:= \lambda^2 \iint d\tau d\tau' \chi(\tau)\chi(\tau') e^{-i(\Omega_A\tau - \Omega_B\tau')} W(x_A(\tau), x_B(\tau')),\label{6} \\
X &:= -\lambda^2 \iint d\tau d\tau' \chi(\tau)\chi(\tau') e^{-i(\Omega_A\tau + \Omega_B\tau')} \Big[ \theta(\tau' - \tau) W(x_A(\tau), x_B(\tau')) \nonumber\\
&\quad + \theta(\tau - \tau') W(x_B(\tau'), x_A(\tau)) \Big].
\label{7}
\end{align}
Here, $W(x,x') = \langle 0|_M \phi(x)\phi(x')|0\rangle_M$ is the Wightman function of the scalar field in the Minkowski vacuum, and $\theta$ denotes the Heaviside step function. The quantities $P_A$ and $P_B$ are the local excitation probabilities of the two detectors, whereas $C$ and $X$ encode the nonlocal correlations generated through their interactions with the quantum field.

We consider two detectors undergoing uniform acceleration with the same proper acceleration $a$ along the $x$ direction. Their trajectories in Minkowski coordinates are given by
\begin{equation}
\begin{aligned}
x_A &= \{ t = a^{-1} \sinh(a\tau_A),\; x = a^{-1} \cosh(a\tau_A),\; y = z = 0 \}, \\
x_B &= \{ t = a^{-1} \sinh(a\tau_B),\; x = a^{-1} \cosh(a\tau_B) + L,\; y = z = 0 \},
\label{8}
\end{aligned}
\end{equation}
where $\tau_A$ and $\tau_B$ are the proper times of detectors $A$ and $B$, respectively, and $L$ characterizes their spatial separation in the chosen Minkowski coordinates. For a massless scalar field in $(3+1)$-dimensional Minkowski spacetime, the Wightman function takes the form \cite{V5}
\begin{equation}
W(x,x') = -\frac{1}{4\pi^2} \frac{1}{(t - t' - i\epsilon)^2 - |\bm{x} - \bm{x}'|^2}.
\label{9}
\end{equation}
Substituting Eqs.~\eqref{8} and \eqref{9} into Eq.~\eqref{5}, the transition probability for detector $D$ is evaluated as \cite{V6,V7}
\begin{equation}
P_D = \frac{\lambda^2 a \sigma}{4\pi^{3/2}} \int_0^\infty d\tilde{s} \, \frac{\cos(2\Omega_D \tilde{s}/a) \, e^{-\tilde{s}^2/(a^2 \sigma^2)} (\sinh^2 \tilde{s} - \tilde{s}^2)}{\tilde{s}^2 \sinh^2 \tilde{s}} + \frac{\lambda^2}{4\pi} \left[ e^{-\Omega_D^2 \sigma^2} - \sqrt{\pi} \Omega_D \sigma \operatorname{Erfc}(\Omega_D \sigma) \right].
\label{10}
\end{equation}
The first term represents the acceleration-dependent contribution to the detector response and vanishes in the inertial limit, whereas the second term is the finite-time response associated with the Gaussian switching and survives even for an inertial detector.

The nonlocal correlation term $X$, which arises from the time-ordered
detector-field interactions, can be written as
\begin{equation}
\begin{split}
X ={} & -\lambda^2 \int_{-\infty}^{\infty} d\tau \int_{-\infty}^{\tau} d\tau' \chi(\tau)\chi(\tau') \Bigg[ e^{-i(\Omega_B \tau + \Omega_A \tau')} W(x_A(\tau'), x_B(\tau)) \\
& + e^{-i(\Omega_A \tau + \Omega_B \tau')} W(x_B(\tau'), x_A(\tau)) \Bigg].
\label{11}
\end{split}
\end{equation}
To facilitate the evaluation of this expression, we introduce the variables
\begin{equation}
u = \tau,\quad s = \tau - \tau'.
\label{12}
\end{equation}
The correlation term thus becomes
\begin{equation}
\begin{split}
X ={} & -\lambda^2 \int_{-\infty}^{\infty} du \int_0^{\infty} ds \Bigg[ e^{(2us - s^2 - 2u^2)/(2\sigma^2) - i(\Omega_A + \Omega_B)u + i\Omega_A s} W(x_A(u-s), x_B(u)) \\
& + e^{(2us - s^2 - 2u^2)/(2\sigma^2) - i(\Omega_A + \Omega_B)u + i\Omega_B s} W(x_B(u-s), x_A(u)) \Bigg].
\label{13}
\end{split}
\end{equation}
Substituting the uniformly accelerated trajectories and the scalar-field
Wightman function, and introducing the energy-gap difference $\Delta\Omega = \Omega_B - \Omega_A$,  
we obtain
\begin{equation}
X = -\frac{\lambda^2 a^2}{4\pi^2} \int_{-\infty}^{\infty} du \int_{0}^{\infty} ds \left[ g_{AB}(u,s) \frac{1}{f_{AB}(u,s)} + g_{BA}(u,s) \frac{1}{f_{BA}(u,s)} \right],
\label{14}
\end{equation}
where the auxiliary functions are defined as
\begin{align}
g_{AB}(u,s) &= \exp\left[ \frac{2us - s^2 - 2u^2}{2\sigma^2} - i(2\Omega_A + \Delta\Omega)u + i\Omega_A s \right],\label{15} \\
f_{AB}(u,s) &= 2 + a^2 L^2 - 2\cosh(as) + 2aL\cosh(au) - 2aL\cosh[a(u-s)] - i\epsilon,\label{16} \\
g_{BA}(u,s) &= \exp\left[ \frac{2us - s^2 - 2u^2}{2\sigma^2} - i(2\Omega_A + \Delta\Omega)u + i\Omega_B s \right],\label{17} \\
f_{BA}(u,s) &= 2 + a^2 L^2 - 2\cosh(as) + 2aL\cosh[a(u-s)] - 2aL\cosh(au) - i\epsilon.\label{18}
\end{align}

The coherence term $C$ can be evaluated in a similar manner using the
variables $(u,s)$ introduced above. Substituting the uniformly accelerated
trajectories and the scalar-field Wightman function into Eq.~\eqref{6}, and
separating the integration domain according to the sign of $s$, we write
\begin{equation}
C = C_1 + C_2,
\label{19}
\end{equation}
where
\begin{equation}
C_1 = -\frac{a^2 \lambda^2}{4\pi^2} \int_{-\infty}^{\infty} du \int_0^{\infty} ds \, e^{-\frac{u^2 + (u-s)^2}{2\sigma^2}}e^{-i[(\Omega_A - \Omega_B)u + \Omega_B s]} \frac{1}{f(u,s) - i\epsilon},
\label{20}
\end{equation}
and
\begin{equation}
C_2 = -\frac{a^2 \lambda^2}{4\pi^2} \int_{-\infty}^{\infty} du \int_0^{\infty} ds \, e^{-\frac{u^2 + (u-s)^2}{2\sigma^2}} e^{i[(\Omega_A - \Omega_B)u + \Omega_B s]}\frac{1}{f(u,s) + i\epsilon}.
\label{21}
\end{equation}
Here, the function $f(u,s)$ is defined as
\begin{equation}
f(u,s) = 2aL \cosh(au) - 2aL \cosh[a(u-s)] + 2 \cosh(as) - 2 - a^2 L^2.
\label{22}
\end{equation}
Applying the Sokhotski formula,
\begin{equation}
\frac{1}{x \pm i\epsilon} = \mathcal{P} \frac{1}{x} \mp i\pi \delta(x),
\label{23}
\end{equation}
and combining the contributions from $C_1$ and $C_2$, the coherence term can finally be expressed as

\begin{equation}
\begin{split}
C ={} & -\frac{a^2 \lambda^2}{2\pi^2} \Bigg\{\mathcal{P} \int_{-\infty}^{\infty} du \int_0^{\infty} ds \, e^{-\frac{s^2 - 2us + 2u^2}{2\sigma^2}} \cos\big[ u(\Omega_A - \Omega_B) + s \Omega_B \big]\frac{1}{f(u,s)} \\
& + \pi \int_{-\infty}^{\infty} du \int_0^{\infty} ds \, e^{-\frac{s^2 - 2us + 2u^2}{2\sigma^2}} \sin\big[ u(\Omega_A - \Omega_B) + s \Omega_B \big] \delta\big[f(u,s) \big]\Bigg\}.
\label{24}
\end{split}
\end{equation}

\section{Directional steering harvesting under the Unruh effect}
Quantum steering is intrinsically directional, and the steerability from
$A$ to $B$ generally differs from that in the reverse direction. To
characterize this directional behavior for the two non-identical detectors,
we employ the analytical steering criteria for two-qubit states~\cite{FFF1,FFF2}. We first consider steering from detector $B$ to
detector $A$ ($B\rightarrow A$). For this direction, we construct the
auxiliary state
\begin{equation}
\tau_{AB} = \frac{1}{\sqrt{3}} \rho_{AB} + \frac{3 - \sqrt{3}}{3} \left( \rho_A \otimes \frac{I}{2} \right),
\label{25}
\end{equation}
where $\rho_A = \operatorname{Tr}_B(\rho_{AB})$ is the reduced state of
detector $A$ and $I$ denotes the $2\times2$ identity matrix. From
Eq.~\eqref{4}, we obtain
\begin{equation}
\rho_A = 
\begin{pmatrix}
1 - P_A & 0 \\
0 & P_A
\end{pmatrix}.
\label{26}
\end{equation}
Substituting Eqs.~\eqref{4} and \eqref{26} into Eq.~\eqref{25} gives
\begin{equation}
\tau_{AB} = 
\begin{pmatrix}
\frac{1}{\sqrt{3}}(1 - P_A - P_B) + f & 0 & 0 & \frac{1}{\sqrt{3}}X \\
0 & \frac{1}{\sqrt{3}}P_B + f & \frac{1}{\sqrt{3}}C & 0 \\
0 & \frac{1}{\sqrt{3}}C^* & \frac{1}{\sqrt{3}}P_A + h & 0 \\
\frac{1}{\sqrt{3}}X^* & 0 & 0 & h
\end{pmatrix},
\label{27}
\end{equation}
where
\begin{equation}
f = \frac{3 - \sqrt{3}}{6}(1 - P_A),\qquad 
h = \frac{3 - \sqrt{3}}{6}P_A.
\label{28}
\end{equation}
According to the steering criterion, the entanglement of
the auxiliary state $\tau_{AB}$ provides a sufficient condition for
$B\rightarrow A$ steering of $\rho_{AB}$. Since $\tau_{AB}$ has an
$X$-state structure, its entanglement can be determined analytically
through concurrence. The condition for $B\rightarrow A$ steering is
therefore satisfied if at least one of the following inequalities holds:
\begin{equation}
|X| > \sqrt{Q_a - Q_b},\quad |C| > \sqrt{Q_c - Q_b},
\label{29}
\end{equation}
where
\begin{align}
Q_a &= \frac{2 + \sqrt{3}}{2} P_A P_B + \frac{1}{4}(1 - P_A - P_B)(P_A + P_B),\label{30} \\
Q_b &= \frac{1}{4}(1 - P_A - P_B)(P_B - P_A),\label{31} \\
Q_c &= \frac{2 - \sqrt{3}}{2} P_A P_B + \frac{1}{4}(1 - P_A - P_B)(P_A + P_B).\label{32}
\end{align}

Similarly, steering from detector $A$ to detector $B$ ($A\rightarrow B$)
can be characterized by introducing the auxiliary state
\begin{equation}
\tau_{BA} = \frac{1}{\sqrt{3}} \rho_{AB} + \frac{3 - \sqrt{3}}{3} \left( \frac{I}{2} \otimes \rho_B \right),
\label{33}
\end{equation}
where $\rho_B=\operatorname{Tr}_A(\rho_{AB})$ is the reduced state of
detector $B$. Following the same procedure as for the reverse direction,
a sufficient condition for $A\rightarrow B$ steering is that at least one
of the following inequalities is satisfied:
\begin{equation}
|X| > \sqrt{Q_a + Q_b}, \qquad |C| > \sqrt{Q_c + Q_b}.
\label{34}
\end{equation}
We therefore quantify the steerability in the two directions by
\begin{equation}
S^{B\rightarrow A} = \max\left\{0,\ |X| - \sqrt{Q_a - Q_b},\ |C| - \sqrt{Q_c - Q_b}\right\},
\label{35}
\end{equation}
and
\begin{equation}
S^{A\rightarrow B} = \max\left\{0,\ |X| - \sqrt{Q_a + Q_b},\ |C| - \sqrt{Q_c + Q_b}\right\}.
\label{36}
\end{equation}

The two steering measures need not be equal, reflecting the intrinsic
directionality of quantum steering. For identical detectors with equal
energy gaps, however, exchange symmetry renders the two steering directions
equivalent, thereby eliminating directional asymmetry. To break this
symmetry, we consider non-identical detectors with
$\Omega_A\neq\Omega_B$. Without loss of generality, we take
$\Omega_B\geq\Omega_A$ and define the energy-gap difference as
$\Delta\Omega=\Omega_B-\Omega_A\geq0$. The resulting directional imbalance
is quantified by the steering asymmetry
\begin{equation}
S^{\Delta}_{AB} = S^{A\rightarrow B} - S^{B\rightarrow A}.
\label{37}
\end{equation}
A nonzero $S^{\Delta}_{AB}$ signals directional asymmetry, with its sign
identifying the favored steering direction and its magnitude quantifying
the degree of asymmetry.
With the analytical framework established, we now turn to a systematic
numerical investigation of directional quantum steering harvesting by
two uniformly accelerated non-identical detectors. We examine how
detector separation, acceleration, and energy-gap difference affect the
strength, asymmetry, and directional regimes of the harvested steering,
with particular emphasis on the directional role of the Unruh effect
and the influence of detector non-identicality.

\begin{figure}[t!]
\begin{minipage}[t]{0.5\linewidth}
\centering
\includegraphics[width=3.0in,height=5.2cm]{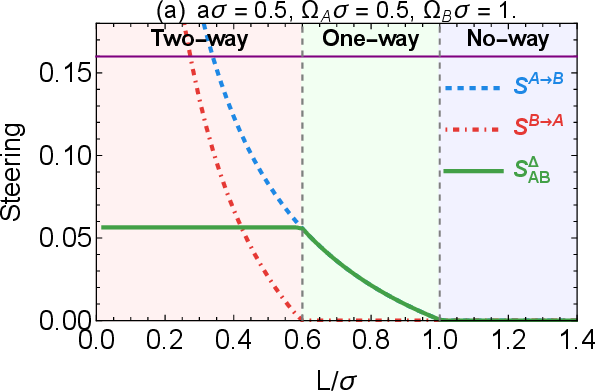}

\label{fig1a}
\end{minipage}%
\begin{minipage}[t]{0.5\linewidth}
\centering
\includegraphics[width=3.0in,height=5.2cm]{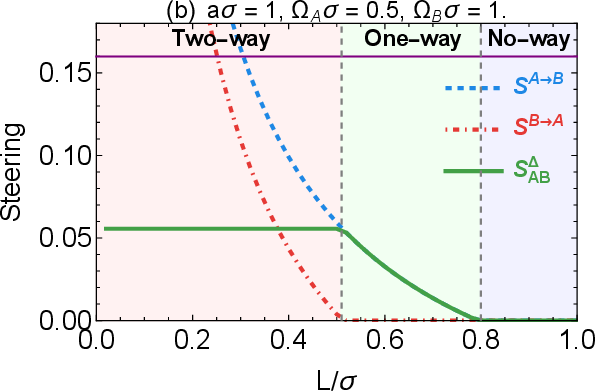}

\label{fig1b}
\end{minipage}
\captionsetup{justification=raggedright,singlelinecheck=false}
\caption{Quantum steering measures $S^{A\rightarrow B}$, $S^{B\rightarrow A}$, and the steering asymmetry $S^{\Delta}_{AB}$ of two uniformly accelerated non-identical detectors as functions of the normalized detector separation $L/\sigma$.}
\label{Fig.1}
\end{figure}

We first investigate the influence of detector separation on directional quantum steering harvesting. Fig.~\ref{Fig.1} presents the steering measures $S^{A\rightarrow B}$ and $S^{B\rightarrow A}$, together with the steering asymmetry $S^{\Delta}_{AB}$, as functions of the normalized detector separation $L/\sigma$ for different accelerations. 
A clear directional asymmetry is observed whenever steering is present, with $S^{A\rightarrow B}$ consistently larger than $S^{B\rightarrow A}$. This asymmetry originates from the non-identical energy gaps of the two detectors. Since the two detectors have different transition energies, their excitation probabilities after interacting with the quantum field are generally different. For the parameters considered here, the smaller-gap detector has a larger excitation probability, i.e., $P_A > P_B$ for $\Omega_A < \Omega_B$. As can be seen from Eq.~(31), this imbalance between the two excitation probabilities leads to a negative value of the asymmetric term, $Q_b < 0$. In the steering measures given by Eqs.~(35) and (36), this negative contribution enters the two directions with opposite signs, thereby modifying the corresponding local terms differently. Consequently, the $A\rightarrow B$ steering condition is less restrictive than the reverse one, resulting in $S^{A\rightarrow B} > S^{B\rightarrow A}$. Before the sudden death of the weaker $B\rightarrow A$ steering, the steering in the two directions
exhibits similar decay behavior, and the difference between them remains nearly unchanged. Therefore, $S^{\Delta}_{AB}$ shows only a weak dependence on $L$ in this region. As the separation further increases, the weaker $B\rightarrow A$ steering undergoes sudden death, leading to the transition from two-way steering to one-way steering. In the one-way steering regime, $S^{\Delta}_{AB}$ is mainly determined by the remaining $A\rightarrow B$ steering and gradually decreases as the latter is further suppressed. Eventually, the sudden death of $A\rightarrow B$ steering leads to the transition from one-way steering to no-way steering, where the steering asymmetry vanishes. This demonstrates that detector separation determines not only the harvested steerability, but also the evolution of the steering asymmetry between the two directions.
\begin{figure}[t!]
\begin{minipage}[t]{0.5\linewidth}
\centering
\includegraphics[width=3.0in,height=5.2cm]{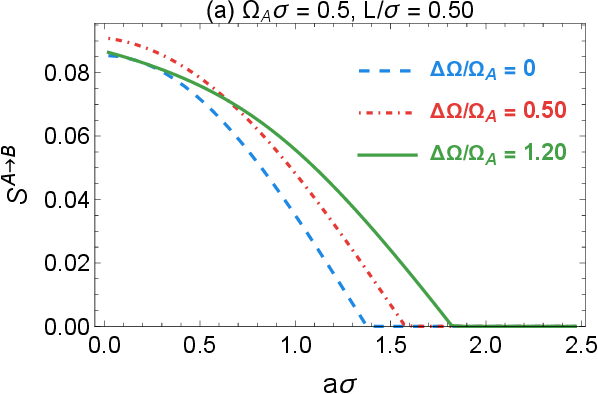}

\label{fig2a}
\end{minipage}%
\begin{minipage}[t]{0.5\linewidth}
\centering
\includegraphics[width=3.0in,height=5.2cm]{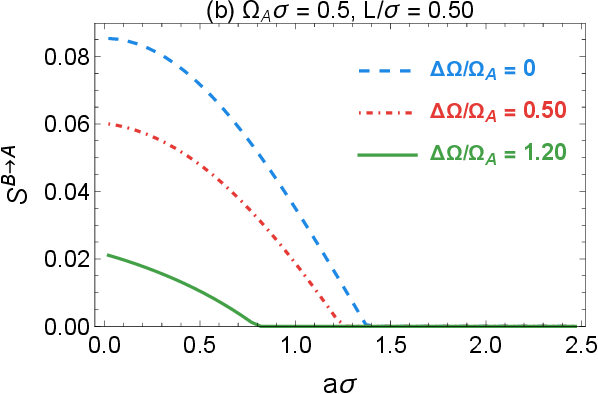}

\label{fig2b}
\end{minipage}

\vspace{0.5cm}

\begin{minipage}[t]{0.5\linewidth}
\centering
\includegraphics[width=3.0in,height=5.2cm]{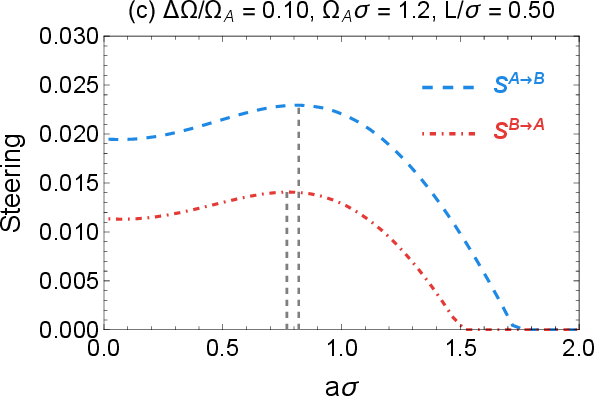}

\label{fig2c}
\end{minipage}%
\begin{minipage}[t]{0.5\linewidth}
\centering
\includegraphics[width=3.0in,height=5.2cm]{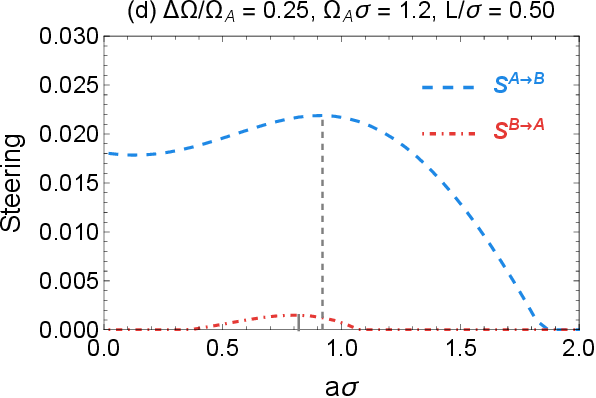}

\label{fig2d}
\end{minipage}
\captionsetup{justification=raggedright,singlelinecheck=false}
\caption{Quantum steering measures $S^{A\rightarrow B}$ and $S^{B\rightarrow A}$ of two uniformly accelerated non-identical detectors as functions of the acceleration $a\sigma$ for different energy-gap differences $\Delta\Omega/\Omega_A$.}
\label{Fig.2}
\end{figure}

We next investigate the influence of the Unruh effect on directional quantum steering harvesting, as shown in Fig.~\ref{Fig.2}. The steering measures $S^{A\rightarrow B}$ and $S^{B\rightarrow A}$ are plotted as functions of the dimensionless acceleration $a\sigma$, with the detector separation fixed at $L/\sigma=0.5$. In contrast to the monotonic suppression with detector separation observed in Fig.~\ref{Fig.1}, the acceleration dependence exhibits qualitatively different behaviors depending on the detector energy scale and energy-gap difference.
For the smaller detector energy scale $\Omega_A\sigma=0.5$, Figs.~\ref{Fig.2}(a) and (b) show that both steering directions decrease monotonically with increasing acceleration and eventually undergo sudden death. The energy-gap difference, however, affects the robustness of the two steering directions in opposite ways. As $\Delta\Omega/\Omega_A$ increases, the sudden-death point of $S^{A\rightarrow B}$ shifts toward larger accelerations, whereas that of $S^{B\rightarrow A}$ shifts toward smaller accelerations. Consequently, increasing the energy-gap difference extends the acceleration range over which $A\rightarrow B$ steering survives while reducing that of the reverse direction. Thus, even when acceleration predominantly suppresses harvested steering, detector non-identicality strongly modifies its directional robustness.

A qualitatively different behavior emerges at the larger detector energy scale $\Omega_A\sigma=1.2$, as shown in Figs.~\ref{Fig.2}(c) and (d). Rather than being monotonically degraded, the harvested steering can initially be enhanced by acceleration, reach a maximum at a finite acceleration, and then decrease as the acceleration is further increased. This nonmonotonic behavior demonstrates that the Unruh effect does not merely act as a source of noise for steering harvesting. Depending on the detector parameters, moderate acceleration can facilitate the extraction of directional steering from the vacuum, whereas sufficiently large acceleration ultimately suppresses it. More importantly, the acceleration-induced response is intrinsically directional. As illustrated in Figs.~\ref{Fig.2}(c) and (d), $S^{A\rightarrow B}$ and $S^{B\rightarrow A}$ attain their maximal values at different accelerations. Consequently, a finite acceleration interval emerges between the two maxima, within which the two steering directions respond oppositely to increasing acceleration: one steering direction is enhanced while the other is simultaneously suppressed. This opposite response demonstrates that the Unruh effect can modify the directional balance of harvested steering rather than simply enhancing or degrading both directions simultaneously.

This directional character becomes particularly evident in Fig.~\ref{Fig.2}(d). At zero acceleration, only $A\rightarrow B$ steering is present. As the acceleration increases, $B\rightarrow A$ steering is activated within an intermediate acceleration interval, driving the system from one-way to two-way steering. With a further increase in acceleration, the activated $B\rightarrow A$ steering reaches its maximum and subsequently undergoes sudden death, restoring one-way steering. Finally, $S^{A\rightarrow B}$ also vanishes at sufficiently large acceleration, resulting in a transition to the no-way steering regime. Therefore, as the acceleration increases, the system successively evolves from one-way to two-way steering, returns to one-way steering, and eventually enters the no-way steering regime. This sequence demonstrates that acceleration can not only enhance or suppress the magnitude of harvested steering, but also dynamically switch its directional regime.

\begin{figure}[t!]
\begin{minipage}[t]{0.5\linewidth}
\centering
\includegraphics[width=3.0in,height=5.2cm]{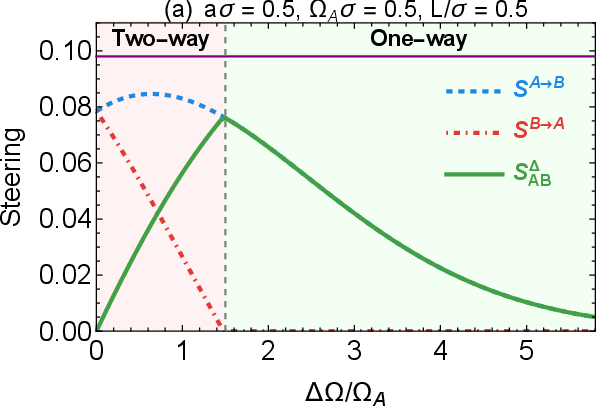}

\label{fig3a}
\end{minipage}%
\begin{minipage}[t]{0.5\linewidth}
\centering
\includegraphics[width=3.0in,height=5.2cm]{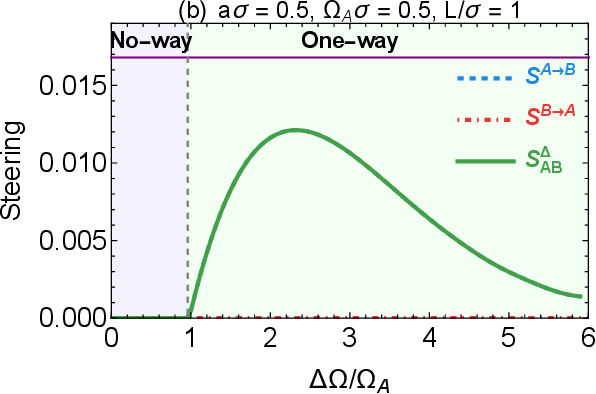}

\label{fig3b}
\end{minipage}

\vspace{0.5cm}

\begin{minipage}[t]{0.5\linewidth}
\centering
\includegraphics[width=3.0in,height=5.2cm]{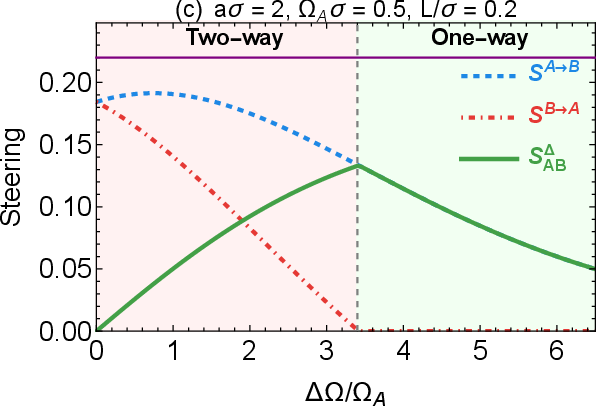}

\label{fig3c}
\end{minipage}%
\begin{minipage}[t]{0.5\linewidth}
\centering
\includegraphics[width=3.0in,height=5.2cm]{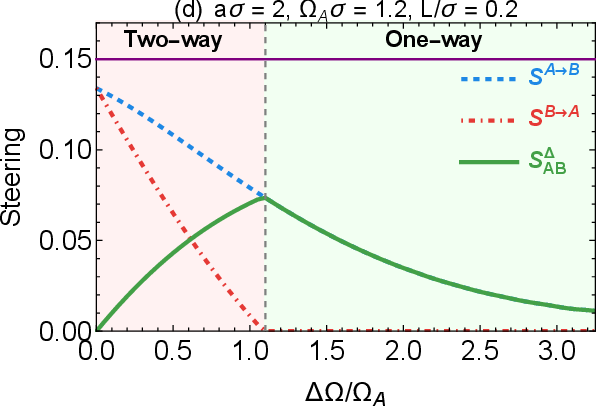}

\label{fig3d}
\end{minipage}
\captionsetup{justification=raggedright,singlelinecheck=false}
\caption{Quantum steering measures $S^{A\rightarrow B}$, $S^{B\rightarrow A}$, and the steering asymmetry $S^{\Delta AB}$ of two uniformly accelerated non-identical detectors as functions of the energy-gap difference $\Delta\Omega/\Omega_A$.}
\label{Fig.3}
\end{figure}

Finally, we examine how detector non-identicality, quantified by the normalized energy-gap difference $\Delta\Omega/\Omega_A$, affects directional quantum steering harvesting. Fig.~\ref{Fig.3} shows the steering measures $S^{A\rightarrow B}$ and $S^{B\rightarrow A}$, together with the steering asymmetry $S^{\Delta}_{AB}$, for different accelerations, detector separations, and detector energy scales.
For weak acceleration and relatively small detector separation, as shown in Fig.~\ref{Fig.3}(a), the two steering directions respond differently to increasing energy-gap difference. The steering $S^{A\rightarrow B}$ first increases, reaches a maximum at an intermediate value of $\Delta\Omega/\Omega_A$, and then decreases, whereas $S^{B\rightarrow A}$ decreases monotonically and eventually undergoes sudden death. Consequently, the system evolves from two-way to one-way steering as the energy-gap mismatch is increased. The steering asymmetry $S^{\Delta}_{AB}$ correspondingly grows, reaches its maximum near the directional-transition point, and then decreases at larger detuning. This behavior shows that a moderate energy-gap mismatch can enhance the directional imbalance of the harvested steering rather than simply suppressing both steering directions.
The role of detector non-identicality becomes even more pronounced when the detector separation is increased. As shown in Fig.~\ref{Fig.3}(b), $S^{B\rightarrow A}$ remains zero throughout the considered parameter range, while $S^{A\rightarrow B}$ is initially absent and becomes nonzero only after $\Delta\Omega/\Omega_A$ exceeds a finite threshold. The energy-gap difference therefore drives a transition from no-way to one-way steering. After being activated, $S^{A\rightarrow B}$ reaches a maximum at an intermediate mismatch and is subsequently suppressed for larger $\Delta\Omega/\Omega_A$. Since the reverse steering remains absent, the steering asymmetry follows the same nonmonotonic behavior as $S^{A\rightarrow B}$. This result demonstrates that detector non-identicality can activate directional steering in a parameter region where steering cannot be harvested for nearly identical detectors.

Figs.~\ref{Fig.3}(c) and~\ref{Fig.3}(d) further demonstrate that the influence of the energy-gap difference depends sensitively on the detector energy scale in the high-acceleration regime. For $\Omega_A\sigma=0.5$, Fig.~\ref{Fig.3}(c) displays behavior qualitatively similar to that in Fig.~\ref{Fig.3}(a): $S^{A\rightarrow B}$ is initially enhanced by increasing $\Delta\Omega/\Omega_A$, whereas $S^{B\rightarrow A}$ decreases continuously and eventually undergoes sudden death, resulting in a transition from two-way to one-way steering. In contrast, when the detector energy scale is increased to $\Omega_A\sigma=1.2$, as shown in Fig.~\ref{Fig.3}(d), the initial enhancement of $S^{A\rightarrow B}$ disappears, and both steering directions decrease monotonically with increasing energy-gap difference. Nevertheless, $S^{B\rightarrow A}$ remains less robust and undergoes sudden death first, again leaving a finite one-way steering regime. An interesting common feature can be identified in Figs.~\ref{Fig.3}(a), \ref{Fig.3}(c), and \ref{Fig.3}(d). In all three cases, the steering asymmetry $S^{\Delta}_{AB}$ reaches its maximum approximately at the point where the weaker $B\rightarrow A$ steering undergoes sudden death. This point therefore marks the boundary between the two-way and one-way steering regimes. Therefore, the maximum of the steering asymmetry provides a clear signature of the transition from two-way to one-way steering.

\section{Conclusion}

In this work, we have investigated directional quantum steering harvesting by two uniformly accelerated non-identical Unruh-DeWitt detectors locally coupled to a massless scalar field in the Minkowski vacuum. By allowing the detectors to possess unequal energy gaps, we explicitly break the exchange symmetry between them and examine how detector separation, acceleration, and energy-gap mismatch jointly govern the strength, asymmetry, and directional regimes of vacuum-harvested steering. Our results show that relativistic acceleration does not simply produce a uniform degradation of steering. Instead, its influence is intrinsically directional and depends sensitively on the detector parameters.

We first find that increasing the detector separation suppresses steering in both directions, but with different robustness. The weaker $B\rightarrow A$ steering undergoes sudden death first, resulting in a transition from two-way to one-way steering, while a further increase in separation eventually eliminates the remaining $A\rightarrow B$ steering and drives the system into the no-way steering regime. The effect of acceleration is considerably richer. At a relatively small detector energy scale, acceleration predominantly suppresses both steering directions. At a larger energy scale, however, moderate acceleration can enhance the harvested steering before the large-acceleration suppression becomes dominant. More importantly, the two steering directions can respond oppositely to acceleration within a finite acceleration interval: one direction is enhanced while the other is simultaneously suppressed. Acceleration can also activate an initially absent $B\rightarrow A$ steering, giving rise to the sequence
\begin{equation}
\text{one-way}
\;\longrightarrow\ \nonumber
\text{two-way}
\;\longrightarrow\;
\text{one-way}
\;\longrightarrow\;
\text{no-way}.
\end{equation}
These results reveal a dual and directional role of the Unruh effect: relativistic acceleration can not only suppress or enhance harvested steering, but also dynamically change its directional structure.
In addition, detector non-identicality provides an additional means of controlling directional steering: a moderate energy-gap mismatch can enhance or even activate $A\rightarrow B$ steering while suppressing the reverse direction, whereas sufficiently large detuning suppresses the harvested steering. Notably, the steering asymmetry $S^{\Delta}_{AB}$ reaches its maximum  when $B\rightarrow A$ steering undergoes sudden death, providing a clear signature of the transition from two-way to one-way steering.

Taken together, our results demonstrate that acceleration and detector non-identicality do more than determine the amount of steering that can be harvested from vacuum fluctuations: they also control which detector can steer the other and induce transitions among different directional steering regimes. In particular, the opposite responses of the two steering directions show that the Unruh effect should not be viewed solely as a source of relativistic degradation, but can instead reshape and, under suitable conditions, enhance the directional structure of harvested quantum correlations. These findings provide a more complete picture of steering harvesting in relativistic settings and suggest that detector parameters may be exploited to control one-way quantum steering in relativistic quantum-information protocols.

\begin{acknowledgments}
This work is supported by the National Natural Science Foundation of China (12575056),   the Natural
Science Foundation of the Liaoning Scientific Committee No. 2026-MS-281, LiaoNing Revitalization Talents Program (XLYC2503099),	Fundamental Research Program of Shanxi Province (202503021212266), and Scientific and Technological Innovation Programs of Higher Education Institutions of Shanxi Province (2025L120).
\end{acknowledgments}

\end{document}